\documentclass[conference]{IEEEtran}
\IEEEoverridecommandlockouts

\usepackage{cite}
\usepackage{amsmath,amssymb,amsfonts}
\usepackage{algorithmic}
\usepackage{graphicx}
\usepackage{textcomp}
\usepackage{xcolor}
\usepackage{multirow}
\usepackage{url}
\usepackage{hyperref}

\usepackage[caption=false,font=footnotesize,labelfont=rm,textfont=rm]{subfig}

\def\BibTeX{{\rm B\kern-.05em{\sc i\kern-.025em b}\kern-.08em
    T\kern-.1667em\lower.7ex\hbox{E}\kern-.125emX}}
\begin{document}

\title{Phoneme-Level Contrastive Learning for User- Defined Keyword Spotting with Flexible Enrollment\\
}

\author{
    \IEEEauthorblockN{
        Kewei Li\IEEEauthorrefmark{1},
        Hengshun Zhou\IEEEauthorrefmark{2},
        Kai Shen\IEEEauthorrefmark{2},
        Yusheng Dai\IEEEauthorrefmark{1},
        Jun Du\IEEEauthorrefmark{1}$^,$\IEEEauthorrefmark{3}, \thanks{\IEEEauthorrefmark{3}Corresponding author}
    }
    \IEEEauthorblockA{\IEEEauthorrefmark{1}NERC-SLIP, University of Science and Technology of China (USTC), Hefei, China}
    \IEEEauthorblockA{\IEEEauthorrefmark{2}iFlytek Research, Hefei, China}
    \IEEEauthorblockA{\{keweili12, zhhs, dalison\}@mail.ustc.edu.cn, kaishen2@iflytek.com, jundu@ustc.edu.cn}
}

\maketitle
\begin{abstract}

User-defined keyword spotting (KWS) enhances the user experience by allowing individuals to customize keywords. However, in open-vocabulary scenarios, most existing methods commonly suffer from high false alarm rates with confusable words and are limited to either audio-only or text-only enrollment. Therefore, in this paper, we first explore the model's robustness against confusable words. Specifically, we propose Phoneme-Level Contrastive Learning (PLCL), which refines and aligns query and source feature representations at the phoneme level. This method enhances the model's disambiguation capability through fine-grained positive and negative comparisons for more accurate alignment, and it is generalizable to jointly optimize both audio-text and audio-audio matching, adapting to various enrollment modes. Furthermore, we maintain a context-agnostic phoneme memory bank to construct confusable negatives for data augmentation. Based on this, a third-category discriminator is specifically designed to distinguish hard negatives. Overall, we develop a robust and flexible KWS system, supporting different modality enrollment methods within a unified framework. Verified on the LibriPhrase dataset, the proposed approach achieves state-of-the-art performance.

\end{abstract}

\section{Introduction}

Keyword Spotting (KWS) is designed to detect user-specified keywords and often serves as the initial trigger for activating speech interactions. Traditional KWS systems predominantly rely on predetermined and static keywords, lacking flexibility for customization
\cite{Guoguochen, Vygon2021LearningER, Tang2017DeepRL, Zhou_Du_Zou_Nian_Lee_Siniscalchi_Watanabe_Scharenborg_Chen_Xiong, Rose1990AHM, LopezEspejo2021DeepSK}. 
In contrast, open-vocabulary KWS adopts a more efficient and flexible paradigm, enabling users to define keywords according to their preferences without retraining or finetuning \cite{Wang2022TripletCF, nishu2024flexible, MetricJung, Li2023AMT, lee2024iphonmatchnet}.

Existing open-vocabulary KWS systems can be categorized based on the enrollment modality into audio enrollment and text enrollment \cite{MatchQueryZhan, MultiAttentionQueryHuang, MultilingualCLLei, Settle2017QuerybyExampleSW}. Audio enrollment is typically implemented using query-by-example methods, where Dynamic Time Warping (DTW) \cite{chen2015query, Fuchs2017SpokenTD, Yusuf2019AnEE} or metric learning \cite{MultilingualqbeReuter, MetricJung} is applied to measure similarity between the reference and incoming audio. Text enrollment, on the other hand, aligns audio and text data within a unified latent space and can mitigate issues caused by low-quality audio compared to audio enrollment\cite{Shin2022LearningAA, MatchingKumari1, lee23d_interspeech}. However, both methods are observed to be sensitive to confusable words that closely resemble the reference keywords in sound, leading to high false trigger rates. Previous approaches using utterance-level contrastive learning\cite{ContrastiveXi} overlook the finer-grained phoneme-level. Recent advancements further highlight the importance of fine-grained modeling at the word or phoneme level to enhance robustness in KWS systems \cite{lee2024iphonmatchnet, ai2024mm}.
For instance, Nishu et al. develop a Phoneme-to-Vector (P2V) database by leveraging connectionist temporal classification (CTC) loss to train audio encoders \cite{nishu2024flexible}. Similarly, Lee et al. utilize a phoneme-level detection loss to enhance robustness across varied pronunciations \cite{lee2024iphonmatchnet}. 
Nevertheless, the phoneme alignment employed in these methods lacks sufficient precision to map phoneme-level audio to text accurately. As a result, the aforementioned constraint methods continue to face challenges, particularly when dealing with confusable words. 
On the other hand, most existing methods are restricted to supporting either audio-only or text-only enrollment, neglecting user convenience and modality fusion for better performance.

To address these challenges, we introduce the Phoneme-Level Contrastive Learning (PLCL) approach. Building on recent advances in contrastive learning
\cite{InternationalNiizumi, wu2018unsupervised, zha2024RNC, oord2018cpc, PretrainingMoummad}, 
PLCL precisely aligns phonemes across both text and audio modalities within a unified representation space through contrastive learning, while also enabling alignment between query and enrolled audio, thereby enhancing audio-audio consistency.
Moreover, we construct a context-agnostic phoneme memory bank alongside a third-category discriminator to generate augmented hard negatives and improve text enrollment. 
Finally, to provide users with maximum flexibility, our verifier flexibly returns results based on the selected enrollment method, including text-only, audio-only, and audio-text modality enrollment options. 
The implementation code are available at the Project page\footnote{\url{https://github.com/wikkk-tp/udkws_FEPLCL}}.
In summary, our contributions are as follows: 1) We propose the PLCL approach and reinforce robust phoneme-level alignment to achieve finer-grained alignment. 2) We achieve a context-agnostic phoneme memory bank to enhance text enrollment and data augmentation. 3) We develop a multimodal flexible enrollment keyword spotting system. 

\begin{figure*}[htbp]
\centerline{\includegraphics[width=7.2in]{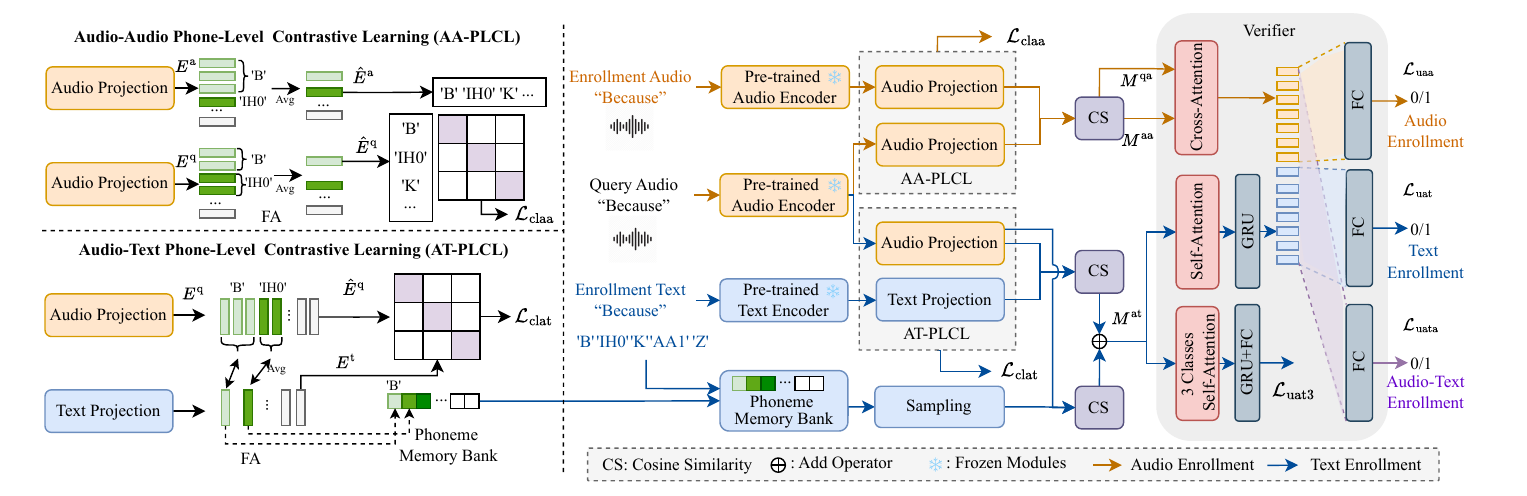}}
\caption{Overall architecture of the proposed model PLCL. The input consists of a query audio paired with either enrollment text or audio, based on the enrollment data, the output is a score used to determine whether the query matches the enrollment word.}
\label{figoverall}
\end{figure*}

\section{Proposed Method}

In this section, we introduce our proposed Phoneme-Level Contrastive Learning (PLCL) system as shown in Fig. \ref{figoverall}. The system consists of audio-text and audio-audio encoders, a phoneme memory bank for data augmentation, and a verifier that generates flexible enrollment results.

\subsection{Phone-Level Contrastive Learning}
\subsubsection{Audio-Text Phone-level Contrastive Learning}

As shown in Fig. \ref{figoverall}, during the training stage, we extract query audio features $E^\mathrm{q} \in \mathbb{R}^{T^\mathrm{q} \times d}$ and text features $E^\mathrm{t} \in \mathbb{R}^{T^\mathrm{t} \times d}$ through their respective encoders and followed by a projection that maps these features into the same embedding space. $T^\mathrm{t}$ and $T^\mathrm{q}$ represent the length of the text and query audio frame, respectively, and $d$ represents the frame dimension.
The projection is trained through contrastive learning using positive paired audio-text data. We use the Montreal Forced Aligner\cite{McAuliffe2017MontrealFA} to perform forced alignment, providing precise audio frame alignment for each phoneme. By applying global averaging on these audio frames, we obtain audio features $\hat{E}^\mathrm{q} \in \mathbb{R}^{\hat{T}^\mathrm{q} \times d}$ that match the length of the text frames.

For a mini-batch, we concatenate all phonemes across the batch to obtain $N$ phoneme samples. Paired audio-text phonemes are treated as positive examples, while all other phonemes within the batch serve as negative examples. This setup allows the model to learn distinct embeddings for phonemes by maximizing the similarity between positive pairs and minimizing it between negative pairs.
Audio-text contrastive learning loss is as InfoNCE loss \cite{Oord2018RepresentationLW, Elizalde2023CLAPLA}, which is defined as follows: 
\begin{equation}
\label{clat}
\mathcal{L}_{\mathrm{clat}} = -\sum_{i=1}^N\log\frac{\exp(\text{sim}(\hat{E}^\mathrm{q}_i, E^\mathrm{t}_i)/\tau)}{\sum_{j=1}^N\exp(\text{sim}(\hat{E}^\mathrm{q}_i, E^\mathrm{t}_j)/\tau)}
\end{equation}
where $\text{sim}(\hat{E}^\mathrm{q}_i, E^\mathrm{t}_i)$ is the cosine similarity between the audio phoneme and the text phoneme, and $i$ represents the specific query example for which the loss is being calculated in the $N$ phoneme samples. $j$ iterates over all possible keys in the mini-batch, including the positive key and all negative keys where $j\neq i$. $\tau$ is the temperature parameter.

\subsubsection{Audio-Audio Phone-level Contrastive Learning}

In the training stage, we also apply the same PLCL method to compare the query audio and enrollment audio. Through the same forced alignment (FA) process described above, we obtain the query audio embedding $\hat{E}^\mathrm{q} \in \mathbb{R}^{\hat{T}^\mathrm{q} \times d}$ and the enrollment audio embedding $E^\mathrm{a} \in \mathbb{R}^{T^\mathrm{a} \times d}$. Then obtain the result of forced alignment $\hat{E}^\mathrm{a} \in \mathbb{R}^{\hat{T}^\mathrm{a} \times d}$, 
$\hat{T}^\mathrm{q}$ and $\hat{T}^\mathrm{a}$ represent the length of query audio and enrollment audio frame, respectively, and are equal to the text length $T^\mathrm{t}$. Audio-audio contrastive learning loss is defined as follows: 
\begin{equation}
\label{claa}
\mathcal{L}_{\mathrm{claa}} = -\sum_{i=1}^N\log\frac{\exp(\text{sim}(\hat{E}^\mathrm{q}_i, \hat{E}^\mathrm{a}_i)/\tau)}{\sum_{j=1}^N\exp(\text{sim}(\hat{E}^\mathrm{q}_i, \hat{E}^\mathrm{a}_j)/\tau)}
\end{equation}

\subsection{Data Augmentation for Hard Negative} \label{DA}

The primary factor influencing model performance is the presence of hard negative examples, which have highly similar pronunciations, differing only in specific phonemes. To address this issue, we construct a set of hard examples and design a classifier specifically for these difficult cases.

We store phoneme embeddings in a memory bank. At each update step, the phoneme embeddings in the memory bank are adjusted using a moving average method \cite{He2019MomentumCF, grill2020bootstrap}, updating the phoneme feature $p_k$ as follows:
\begin{equation}
\label{MA}
p_k\leftarrow \alpha p_k+(1-\alpha)p^{\mathrm{new}}_k
\end{equation}
here, $p_k$ is the feature representation of the $k$-th phoneme.
the new embedding $p^{\mathrm{new}}_k$ is combined with the existing memory $p_k$ using a momentum coefficient $\alpha$ set to 0.8. Additionally, we selectively average high-quality phoneme data with the memory bank to enhance the overall representation.

To enhance recognition for hard negatives, we randomly extract phonemes from the independent phoneme memory bank and insert, replace, or delete phonemes in positive examples to create hard negative examples. Inspired by \cite{sfx2024icme}, we treated the training set and the generated and original hard negative examples as the third category, using a 3-class discriminator to assist in the training.

Additionally, we search for similar words within the training set and increase the number of pairs involving less accurately recognized examples based on the training results. These words are then rearranged and paired with existing words from the training set.

\subsection{Multi-modal Flexible Enrollment KWS}
In user-defined KWS, users can enroll using either audio or text. 
When a query audio is input, the verifier outputs a score indicating whether it matches the enrolled keyword. 
In the inference stage, the system is activated based on the user's enrollment modality. If input from a particular modality is missing, it is masked appropriately. 
In the training stage, we train with inputs from all enrollment modalities. Below, we present our training process.

\subsubsection{Text Enrollment KWS}
When the user enrolls with text, we use the fine-grained constraints of PLCL to map text and speech to the same space and achieve fine-grained alignment,
audio and text are automatically aligned without the need for forced alignment. A similarity matrix is then generated using cosine similarity. Context-agnostic phonemes are sampled from the memory bank and added to this matrix. Finally, we apply self-attention to the similarity matrix $M^\mathrm{at} \in  \mathbb{R}^{T^\mathrm{t} \times T^\mathrm{q}}$.
\begin{equation}
\label{att}
E^\mathrm{at}=\text{Self-Attention}(M^\mathrm{at},M^\mathrm{at},M^\mathrm{at})
\end{equation}

The final output is processed through a GRU and a fully connected (FC) layer to obtain the posterior score, representing the degree of alignment between the text and the audio at the utterance level.
The text enrollment part can be composed of utterance level binary cross-entropy (BCE) loss, denoted as $\mathcal{L}_{\mathrm{uat}}$, data augmentation cross-entropy loss (CE) for 3 classes classification, denoted as $\mathcal{L}_{\mathrm{uat3}}$,
and the aforementioned loss $\mathcal{L}_{\mathrm{clat}}$, the final text enrollment is optimized using the following loss function:
\begin{equation}
\mathcal{L}_{\mathrm{at}} = \mathcal{L}_{\mathrm{uat}} + \mathcal{L}_{\mathrm{uat3}} + \mathcal{L}_{\mathrm{clat}}
\end{equation}

\subsubsection{Audio Enrollment KWS}
When the user enrolls with audio, we compute the cosine similarity of the features from the projection output, after which we obtain the similarity matrix $M^\mathrm{aa} \in \mathbb{R}^{T^\mathrm{a} \times T^\mathrm{q}}$.
We take the maximum value along the query audio dimension to obtain $M^\mathrm{qa} \in  \mathbb{R}^{T^\mathrm{a}}$, which is then used as the query. $M^\mathrm{aa}$ is used as the key and value in the cross-attention mechanism.
\begin{equation}
\label{aac}
E^\mathrm{aa}=\text{Cross-Attention}(M^\mathrm{qa},M^\mathrm{aa},M^\mathrm{aa})
\end{equation}

The audio enrollment part can be composed of utterance-level focal loss\cite{lin2017focal}, denoted as $\mathcal{L}_{\mathrm{uaa}}$ and the aforementioned loss $\mathcal{L}_{\mathrm{claa}}$. The final audio enrollment is optimized using the following loss function:
\begin{equation}
\mathcal{L}_{\mathrm{aa}} = \mathcal{L}_{\mathrm{uaa}} + \mathcal{L}_{\mathrm{claa}}
\end{equation}

\subsubsection{Audio-Text Enrollment KWS}

\begin{table}[htbp]
\caption{Evaluation of PLCL model on Libriphrase Hard ($\textbf{LP}_\textbf{H}$) and Libriphrase Easy ($\textbf{LP}_\textbf{E}$). ``*'' indicates a system that uses data augmentation. ``EM'' represents the enrollment method, ``T'' represents text enrollment, ``A'' represents audio enrollment, ``TA'' represents text and audio enrollment.
}
\begin{center}
\setlength{\tabcolsep}{2.3mm}
\renewcommand{\arraystretch}{1.1}
\begin{tabular}{llccccc}

\hline
\multirow{2}{*}{{\begin{tabular}[c]{@{}c@{}}\textbf{EM}\end{tabular}
}} & \multirow{2}{*}{\textbf{Method}} &  \multicolumn{2}{c}{\textbf{AUC(\%)}$\uparrow$}        & \multicolumn{2}{c}{\textbf{EER(\%)}$\downarrow$}            \\ \cline{3-6} 
 &    & $\textbf{LP}_\textbf{H}$ & $\textbf{LP}_\textbf{E}$  & $\textbf{LP}_\textbf{H}$ & $\textbf{LP}_\textbf{E}$ \\ \hline
T &  CMCD\cite{Shin2022LearningAA}  & 73.58 & 96.70  & 32.90 & 8.42   \\  
T &  CLAD\cite{ContrastiveXi}& 76.15 &97.03 &30.30&8.65   \\
T &  iPhonMatchNet\cite{lee2024iphonmatchnet}& 88.23 &99.59 &19.70&2.40\\
T &  CED*\cite{nishu2024flexible}& 92.70 &99.84 &14.40&1.70   \\
T &  AdaKWS-Tiny*\cite{Navon2023OpenVocabularyKW}& 93.75 &99.80 &13.47&1.61   \\
T &  AdaKWS-Small*\cite{Navon2023OpenVocabularyKW}& 95.09 &99.82 &11.48&1.21  \\
TA &  MM-KWS*\cite{ai2024mm}& \underline{96.25} & \underline{99.95} &\underline{9.30}& \underline{0.68}   \\ \hline
TA 
&  \textbf{PLCL}& 95.06 &99.91 &11.08&1.02   \\
A &\multicolumn{1}{l}{\textbf{PLCL}*} &93.09&99.83 & 13.94&1.80  \\
T &\multicolumn{1}{l}{\textbf{PLCL}*} &95.56&99.91  &9.96&1.21   \\
TA &  \textbf{PLCL}*& \textbf{96.59}&\textbf{99.97} &\textbf{8.47}&\textbf{0.57}\\
 \hline
\end{tabular}
\label{sota}
\end{center}
\end{table}

When both text and audio enrollment are used, we concatenate the features from both and apply a FC layer to obtain the final output. 
The loss is composed of an utterance-level Binary Cross-Entropy (BCE) loss $\mathcal{L}_{\mathrm{uata}}$. The PLCL system can be optimized using the following loss criteria:
\begin{equation}
\label{a-t}
\mathcal{L} = \mathcal{L}_{\mathrm{at}} + \mathcal{L}_{\mathrm{aa}} + \mathcal{L}_{\mathrm{uata}}
\end{equation}


\section{Experiment Configuration}
\subsection{Datasets}

We utilize the LibriPhrase \cite{Shin2022LearningAA} dataset, which consists of phrases extracted from the LibriSpeech \cite{LibrispeechPanayotov} corpus, specifically from the train-clean-100 and train-clean-360 subsets, containing phrases ranging from 1 to 4 words. The evaluation set was derived from the train-others-500 subset, and the negative examples were categorized into two groups, easy ($\textbf{LP}_\textbf{E}$) and hard ($\textbf{LP}_\textbf{H}$), according to their Levenshtein distances \cite{Levenshtein1965BinaryCC}.
The anchor text was used for text enrollment, while the anchor audio was used for audio enrollment.

\subsection{Implementation Details}

We utilize the pre-trained Whisper-Tiny encoder \cite{radford2023robust} as the audio encoder.
The text encoder includes a pre-trained grapheme-to-phoneme (G2P) model\cite{g2pE2019}.
The projection layer consists of a LayerNorm, followed by a fully connected (FC) layer and a non-linear activation function. This sequence ensures that both the text and audio embeddings are mapped to a 128-dimensional space.

We train the model for 50 epochs on four V100 GPUs, using the SGD optimizer to minimize the loss with an initial learning rate of 0.01 and a batch size of 64. 

\begin{figure}[htbp]
\centerline{\includegraphics[width=3.6in]{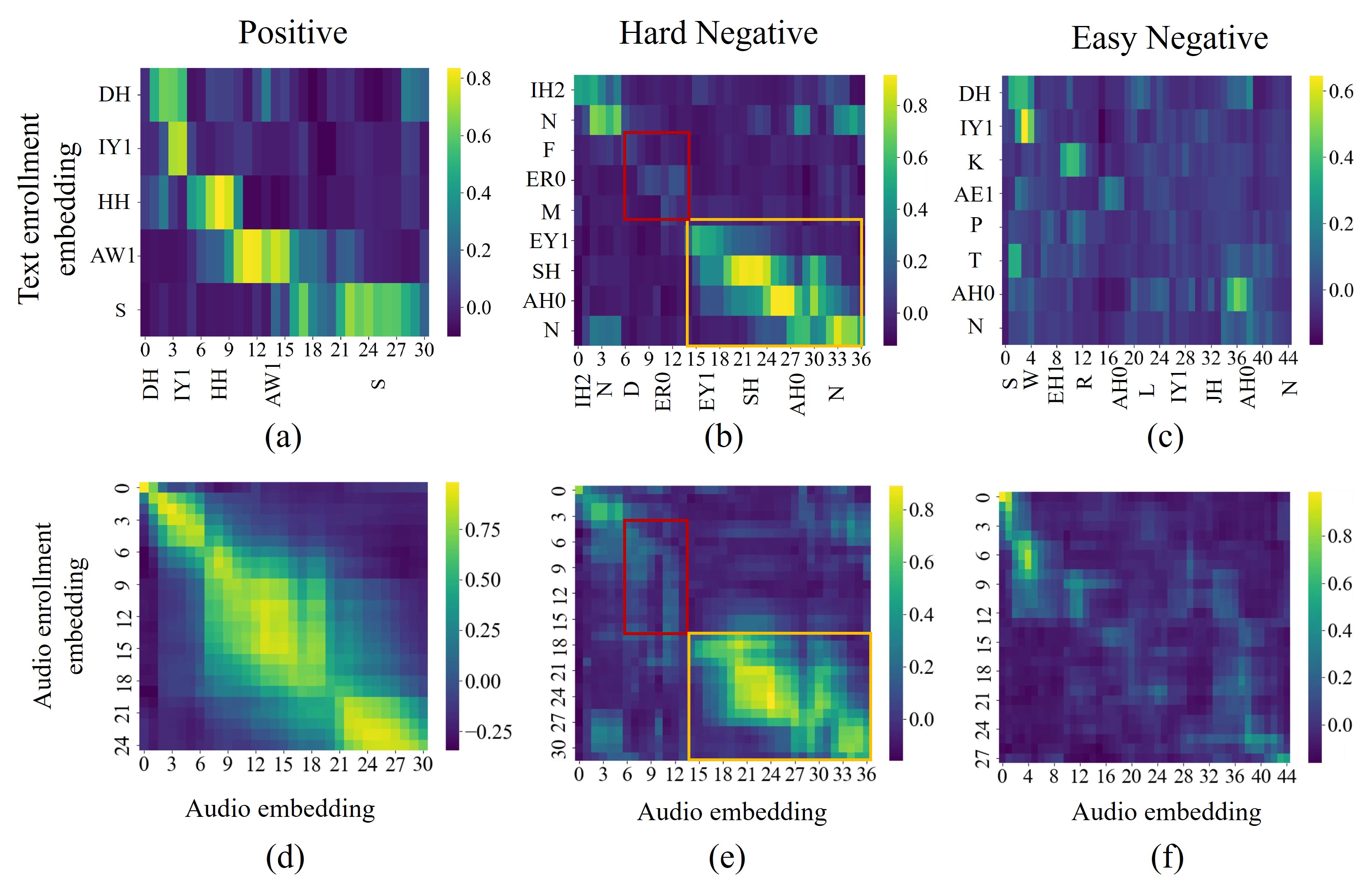}}
\caption{Visualization of attention maps. The positive examples (a) and (d) target keyword is ``the house'',  the hard negative examples (b) and (e) target keyword is ``information'' while the query audio is ``induration'', the easy negative examples (c) and (f) target keyword is ``the captain'' while the query audio is ``swear allegiance''.}
\label{cs_m}
\end{figure}

\section{Results and Analysis}

\subsection{Comparative Evaluation of PLCL}
As shown in Table \ref{sota}, our proposed method PLCL, not only provides a more flexible enrollment approach compared to other methods, but also achieves an Area Under the ROC Curve (AUC) of 99.97\% and an Equal-Error-Rate (EER) of 0.57\% on the $\textbf{LP}_\textbf{E}$ dataset, and an AUC of 96.59\% and an EER of 8.47\% on the $\textbf{LP}_\textbf{H}$ dataset, these are the results after model fusion. Prior to fusion, we were still able to achieve an AUC of 96.41\% and EER of 8.85\% on the $\textbf{LP}_\textbf{H}$, surpassing current state-of-the-art systems. This performance significantly outperforms utterance-level contrastive learning method \cite{ContrastiveXi}.
Compared to the method in \cite{ai2024mm}, which generates 1.5 million paired training data samples, our approach, trained on the original dataset, achieves better performance by enforcing phoneme-level alignment. 

\subsection{Visual Analysis}

Fig. \ref{cs_m} illustrates the attention maps for cosine similarity calculations between audio-text and audio-audio embeddings, specifically analyzing positive, hard negative, and easy negative examples.
The figure highlights that hard negatives, due to their numerous phoneme similarities with positives, are more likely to be misclassified. However, our model can distinguish these hard examples through phoneme alignment.

\begin{figure}[htbp]
\centerline{\includegraphics[width=3.2in]{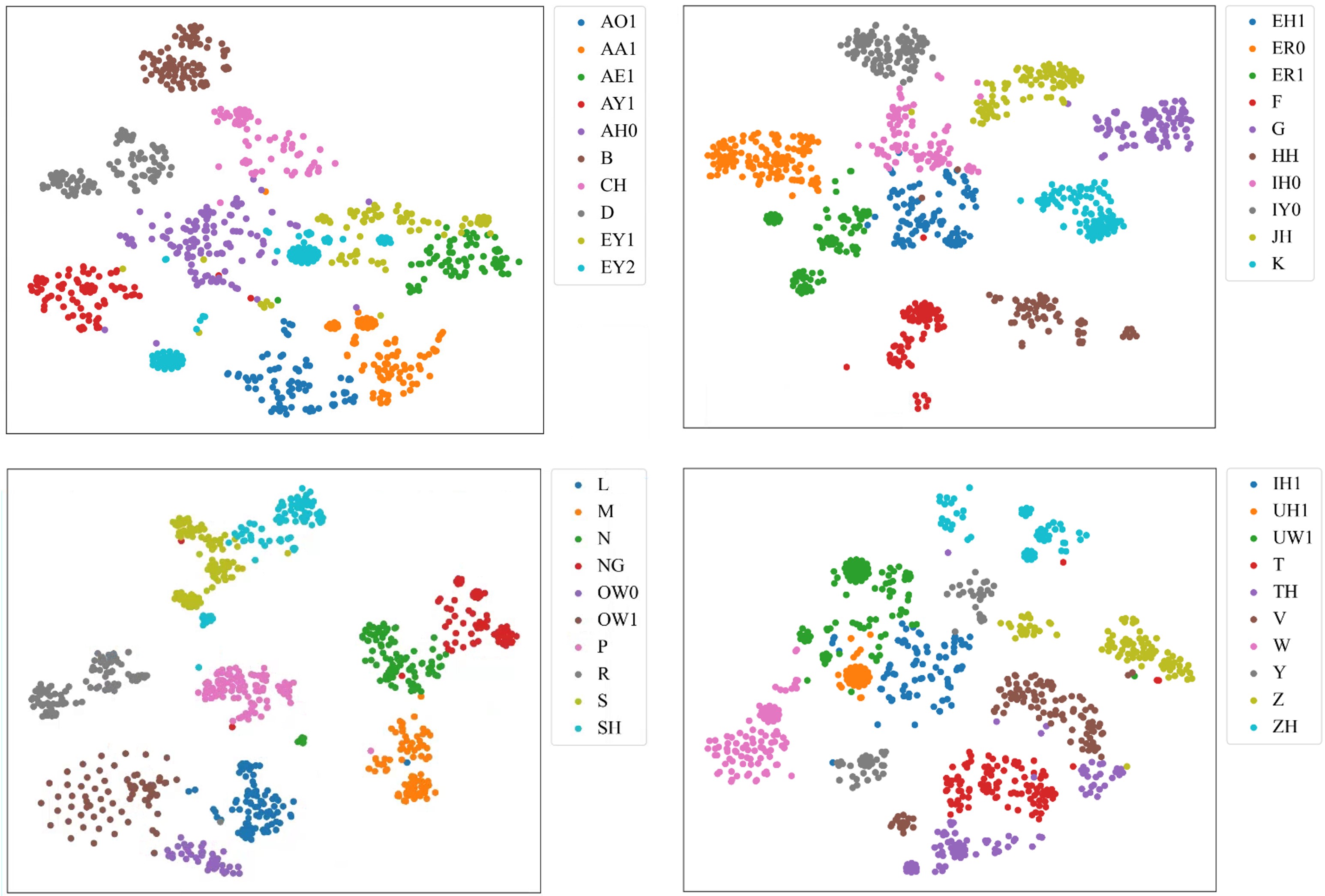}}
\caption{Visualization of t-SNE for various phonemes.}
\label{tsne}
\end{figure}

In Fig. \ref{tsne}, we randomly sample embeddings of different phonemes obtained after contrastive learning from the projection and visualize their t-SNE distribution. The figure shows clear intra-phoneme clustering and separation between different phonemes. 
We analyze challenging cases like ``the prince’s'' and ``the princes'', which have identical pronunciations, 
Additionally, there are cases where only a single phoneme differs, and this phoneme is also closely clustered in the t-SNE plot, such as ``S'' and ``SH'', both of which can lead to misclassifications due to their similarities.

\subsection{Ablation Studies}

Table \ref{Ablationtext} presents the results of our ablation study on text enrollment. The AUC performance on $\textbf{LP}_\textbf{H}$ drops from 92.45\% to 86.71\% when phoneme-level contrastive learning is removed, emphasizing the importance and effectiveness of incorporating this method in our approach. Additionally, data augmentation further improves our system by 3.11\%, highlighting the importance of handling challenging examples.

\begin{table}[htbp]
\caption{Ablation studies of text enrollment}
\begin{center}
\setlength{\tabcolsep}{2.5mm}
\renewcommand{\arraystretch}{1.1}
\begin{tabular}{ccccc}

\hline
\multirow{2}{*}{\textbf{Method}} &  \multicolumn{2}{c}{\textbf{AUC(\%)}$\uparrow$}             & \multicolumn{2}{c}{\textbf{EER(\%)}$\downarrow$}            \\ \cline{2-5} 
      & $\mathrm{\textbf{LP}}_\mathrm{H}$ & $\textbf{LP}_\textbf{E}$ & $\textbf{LP}_\textbf{H}$ & $\textbf{LP}_\textbf{E}$  \\ \hline
\multicolumn{1}{l}{AT-PLCL*} &\textbf{95.56}&\textbf{99.91}  &\textbf{9.96}&\textbf{1.21} \\
\multicolumn{1}{l}{AT-PLCL} & 92.45&99.81&14.08&1.71   \\
\multicolumn{1}{l}{\hspace{2em}w/o memory bank} &      92.36&99.72&14.29&2.06   \\
\multicolumn{1}{l}{\hspace{2em}w/o phoneme-level} &    86.71&98.49&21.05&5.81   \\
\hline

\end{tabular}
\label{Ablationtext}
\end{center}
\end{table}

Table \ref{Ablationaudio} presents the results of our ablation study on audio enrollment. It shows that incorporating phoneme-level contrastive learning leads to an absolute 2.13\% improvement. 
Additionally, using both audio and text for enrollment outperforms single-modality enrollment, indicating that audio provides a complementary benefit to text.

\begin{table}[!ht]
\caption{Ablation studies of audio enrollment}
\begin{center}
\setlength{\tabcolsep}{3mm}
\renewcommand{\arraystretch}{1.1}
\begin{tabular}{ccccc}

\hline
\multirow{2}{*}{\textbf{Method}} &  \multicolumn{2}{c}{\textbf{AUC(\%)}$\uparrow$}             & \multicolumn{2}{c}{\textbf{EER(\%)}$\downarrow$}            \\ \cline{2-5} 
      & $\textbf{LP}_\textbf{H}$ & $\textbf{LP}_\textbf{E}$ & $\textbf{LP}_\textbf{H}$ & $\textbf{LP}_\textbf{E}$  \\ \hline
\multicolumn{1}{l}{PLCL*} & \textbf{96.59}&\textbf{99.97} &\textbf{8.47}&\textbf{0.57} \\
\multicolumn{1}{l}{AA-PLCL*}  &93.09&99.83 & 13.94&1.80    \\
\multicolumn{1}{l}{AA-PLCL} & 92.95&99.82&14.15&1.87   \\
\multicolumn{1}{l}{\hspace{2em}w/o phoneme-level} &    90.82&99.80&16.59&1.88   \\
\hline

\end{tabular}
\label{Ablationaudio}
\end{center}
\end{table}

\section{Conclusion}
In this paper, we introduce PLCL, an innovative approach for user-defined keyword spotting, empowering users with flexible keyword enrollment. By applying contrastive learning, we ensure precise feature representations at the phoneme-level, aligning audio-text and audio-audio pairs for more accurate alignment. With a context-agnostic phoneme memory bank and a third-category discriminator data augmentation for confusable keywords. The system's performance and robustness are further enhanced. 
By fusing text and audio enrollment, the proposed approach overcomes single-modality limitations. 
Experiments prove that the proposed PLCL approach achieves state-of-the-art performance.

\bibliographystyle{IEEEtran}
\bibliography{mybib2}

\end{document}